# Accelerating the calculation of electron-phonon coupling by machine learning methods

Yang Zhong[1,2], Zhiguo Tao[1,2], Weibin Chu[1,2], Xingao Gong[1,2], Hongjun Xiang[1,2*]

[1]Key Laboratory of Computational Physical Sciences (Ministry of Education), Institute of Computational Physical Sciences, State Key Laboratory of Surface Physics, and Department of Physics, Fudan University, Shanghai, 200433, China

[2]Shanghai Qi Zhi Institute, Shanghai, 200030, China

*E-mail: hxiang@fudan.edu.cn

## Abstract

Electron-phonon coupling (EPC) plays an important role in many fundamental physical phenomena, but the high computational cost of the EPC matrix hinders the theoretical research on them. In this paper, an analytical formula is derived to calculate the EPC matrix in terms of the Hamiltonian and its gradient in the nonorthogonal atomic orbital bases. The recently-developed E(3) equivariant neural network is used to directly predict the Hamiltonian and its gradient needed by the formula, thus bypassing the expensive self-consistent iterations in DFT. The correctness of the proposed EPC calculation formula and the accuracy of the predicted EPC values of the network are illustrated by the tests on a water molecule and a $MoS_2$ crystal.

## Introduction

Phonon, one of the fundamental quasiparticles in periodic solids, is generated by the collective vibrations of the nucleus[1]. The electron-phonon coupling (EPC) is ubiquitous in solid materials and plays an important role in a variety of physical phenomena[2, 3]. In metals and semiconductor materials, low-energy electronic excitations caused by phonons can significantly affect their transport and thermodynamic properties[3-5]. The light absorption of indirect band gap semiconductors is promoted by the momentum exchange with the phonons. In Bardeen-Cooper-Schrieffer (BCS) superconductors[6], EPC can provide the weak attraction between electrons to form Cooper electron pairs, resulting in superconducting quantum phase transitions. Although EPC is closely related to many fundamental physical properties of the materials, it is still a challenge to compute the EPC matrix accurately due to the high computational cost. Especially, when we use HSE functional or spin-orbit coupling (SOC) correction, the calculation amount of the EPC matrix will be prohibitive.

EPC matrix can be directly calculated by the density functional perturbation theory (DFPT)[7], which requires high computational complexity and scales poorly as the system size increases. In recent years, the method of calculating the EPC matrix based on the Wannier basis has been widely used[8-11], like the well-known EPW[10] and Perturbo[8] packages. The local Wannier functions in the real space can be transformed into the Bloch wavefunctions in the reciprocal space by the Fourier transformation, which can reduce the calculation amount of DFT in the reciprocal space. Nevertheless, these DFT methods still consume a lot of computing resources to compute the EPC matrix.

The development of machine learning (ML) in condensed matter physics provides a shortcut that can bypass the high computational cost of DFT[12-14]. The kernel ridge regression (KRR) and Gaussian Process Regression (GPR) methods were first used to learn the tight-binding Hamiltonian[15-17]. To achieve better multi-element generalization ability, some Hamiltonian prediction models based on message-passing neural networks (MPNNs) were then proposed and achieved higher accuracy than traditional kernel-based methods[18-21]. Recently, E(3) equivariant MPNNs have been used to build a direct mapping from the structures to the ab initio

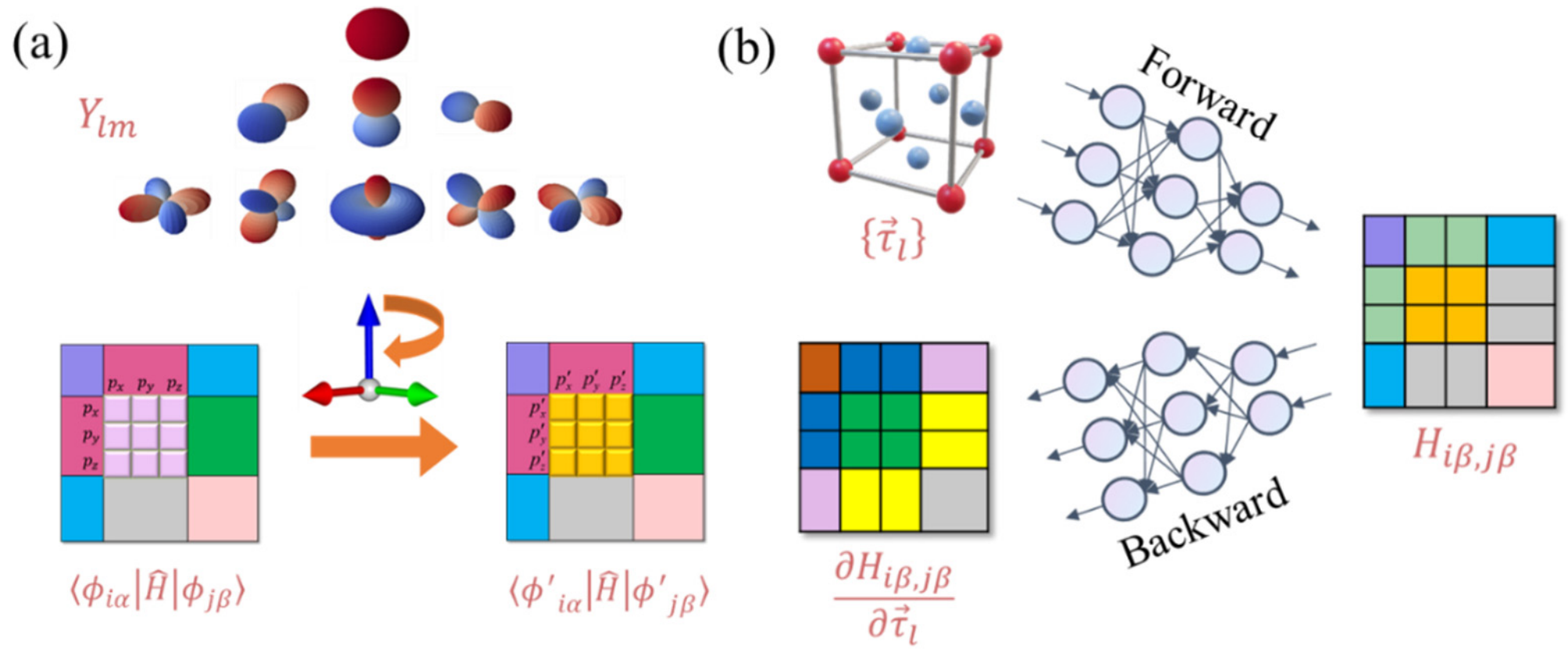

**Fig. 1**. The schematic diagrams of the equivariance of the Hamiltonian matrix and the prediction of the derivatives of the Hamiltonian matrix elements. (a) The schematic diagram of the equivariance of the Hamiltonian matrix. Because of the spherical harmonic part of the atomic orbital basis, the tight-binding Hamiltonian matrix transforms equivariantly with the rotation of the coordinate system. (b) The illustration of the prediction of the derivatives of the Hamiltonian matrix elements. The forward propagation of the neural network can establish an equivariant mapping from atomic coordinates to the Hamiltonian matrix. The gradients of the Hamiltonian matrix elements with respect to each atomic coordinate can be obtained at the input side through the backward propagation of the gradient layer by layer.

Hamiltonian matrices and have shown state-of-the-art accuracy and generalization ability[22, 23]. Different from the invariant scalar properties such as energy and bandgap, the Hamiltonian matrix transforms equivariantly under rotation (see Fig. 1(a)), spatial inversion, and time-reversal operations. The EPC matrix, as the response of the Hamiltonian matrix to the phonon perturbations, should also fulfill all the equivariance of the Hamiltonian matrix. Therefore, We develop the EPC prediction framework based on the E(3) equivariant HamGNN (previously called ‘HamNet’) model proposed in our previous work[22]. Since the Hamiltonian matrix can be analytically represented as a nonlinear function of the atomic coordinates by the deep neural networks, the derivatives of the Hamiltonian elements with respect to the atomic coordinates can be calculated by the backpropagation of the networks (see Fig. 1(b)). In addition, the gradients of the Hamiltonian elements obtained by the backpropagation of the E(3) equivariant

MPNNs automatically satisfy all the fundamental equivariance of the Hamiltonian matrix as well. When obtaining the Hamiltonian and its derivative, the EPC matrix can be calculated by combining the information of lattice vibrations.

In this work, an analytical formula of the EPC matrix in the non-orthogonal atomic orbital bases is derived and expressed as a function of the Hamiltonian matrix and its gradient with respect to the atomic coordinates. The E(3) equivariant HamGNN network is used to predict the Hamiltonian matrix and its gradient needed by the formula. The tests on a water molecule and a $MoS_2$ supercell show that our predicted EPC matrix is very close to the value calculated by DFT, indicating that this is a promising efficient method for calculating the EPC matrix.

# Results

### *The formula of the EPC matrix in the non-orthogonal atomic orbital bases*

The EPC matrix is defined as[2, 9]:

$$g\left(\mu\vec{k}_1,\nu\vec{k}_2,b\vec{q}\right)=\left\langle \psi_\mu^{\vec{k}_1}\left|\frac{\partial\hat{H}}{\partial\eta_b^{\vec{q}}}\right|\psi_\nu^{\vec{k}_2}\right\rangle, \tag{1}$$

where $\psi_\mu^{\vec{k}_1}$ is the Bloch wave function with band index $\mu$ and crystal momentum $\vec{k}_1$ in the Brillouin region, $\psi_\upsilon^{\vec{k}_2}$ is the Bloch wave function with band index $\upsilon$ and crystal momentum $\vec{k}_2$ in the Brillouin region, H is the Hamiltonian of the system, and $\eta_b^{\vec{q}}$ is the normal coordinates for the phonon mode with phonon band index *b* and crystal momentum $\vec{q}$ in Brillouin zone. Supercells are often used in the calculation of lattice vibrations. Here we consider a supercell (denoted by $\mathrm{Cell}_p$) containing $N_p$ unit cells and denote the position of each cell as $\vec{R}_p$ ($p=1,2,\cdots,N_p$). Each cell contains *M* atoms whose Cartesian coordinates are $\tau_{l_\sigma,\vec{R}_p}=\tau^0_{l_\sigma,\vec{R}_p}+\Delta\tau_{l_\sigma,\vec{R}_p}$ ($l$ = 1, 2, …, *M*), where $\tau^0_{l_\sigma,\vec{R}_p}$ is the coordinate of the equilibrium position of atom $l$ in the $p$th supercell, $\Delta\tau_{l_\sigma,\vec{R}_p}$ denotes the displacement from the equilibrium position, σ denotes the $x$, $y$, and $z$ directions.

In harmonic approximation, the atomic vibration can be expressed in the normal coordinates:

$$\Delta\tau_{l_\sigma,\vec{R}_p} = \frac{1}{\sqrt{M_l}} \sum_{b\vec{q}} \xi_{l_\sigma}^{b\vec{q}} \eta_b^{\vec{q}} e^{i\vec{q}\cdot\vec{R}_p}, \tag{2}$$

where $\xi_{l_\sigma}^{b\vec{q}}$ is the eigenvector of the phonon mode with band index $b$ and phonon crystal momentum $\vec{q}$. According to Eq. (2), the EPC matrix can be expressed as:

$$g\left(\mu\vec{k}_1, \nu\vec{k}_2, b\vec{q}\right) = \sum_{l\sigma p} \frac{\xi_{l_\sigma}^{b\vec{q}} e^{i\vec{q}\cdot\vec{R}_p}}{\sqrt{M_l}} \left\langle \psi_\mu^{\vec{k}_1} \middle| \frac{\partial\hat{H}}{\partial\tau_{l_\sigma,\vec{R}_p}} \middle| \psi_\nu^{\vec{k}_2} \right\rangle. \tag{3}$$

The core of the calculation of the EPC matrix is to calculate the matrix $\left\langle \psi_\mu^{\vec{k}_1} \middle| \frac{\partial\hat{H}}{\partial\tau_{l_\sigma,\vec{R}_p}} \middle| \psi_\nu^{\vec{k}_2} \right\rangle$, the expectation value of the gradient operator of Hamiltonian. In many cases involving large systems, the calculation of the EPC matrix only needs to consider the Γ point phonon (i.e., $\vec{q} = 0$). In the following discussion, we will consider only the EPC matrix at Γ points and derive the formula of $\left\langle \psi_\mu^{\vec{k}} \middle| \frac{\partial\hat{H}}{\partial\tau_{l_\sigma,\vec{R}_p}} \middle| \psi_\nu^{\vec{k}} \right\rangle$ in the non-orthogonal atomic orbital bases. To make the formula more concise, we omit the subscript index $\vec{R}_p$ in $\vec{\tau}_{l,\vec{R}_p}$ and use $\vec{\tau}_l$ instead. In the non-orthogonal atomic orbital bases, the Hamiltonian operator of Cell$_p$ can be expressed as:

$$\hat{H}^{\vec{k}} = \sum_{i'j'\alpha'\beta'} \tilde{H}_{i'\alpha',j'\beta'}^{\vec{k}} \left| \varphi_{i'\alpha'}^{\vec{k}} \right\rangle \left\langle \varphi_{j'\beta'}^{\vec{k}} \right|, \tag{4}$$

where $\varphi_{i'\alpha'}^{\vec{k}} = \frac{1}{\sqrt{N}} \sum_{m'}^{N_e} e^{i\vec{R}_{m'}\cdot\vec{k}} \phi_{i'\alpha'}\left(\vec{r} - \vec{\tau}_{i'} - \vec{R}'_{m'}\right)$ and $\varphi_{j'\beta'}^{\vec{k}} = \frac{1}{\sqrt{N}} \sum_{n'}^{N_e} e^{i\vec{R}_{n'}\cdot\vec{k}} \phi_{j'\beta'}\left(\vec{r} - \vec{\tau}_{j'} - \vec{R}'_{n'}\right)$ are the crystal-periodic basis functions from the Fourier transform of the real-space atomic orbits $\phi_{i'\alpha'}$ and $\phi_{j'\beta'}$. $\alpha' \equiv n'_i l'_i m'_i$ and $\beta' \equiv n'_j l'_j m'_j$ denote the quantum numbers. According to $\left\langle \varphi_{i\alpha}^{\vec{k}} \middle| \hat{H}^{\vec{k}} \middle| \varphi_{j\beta}^{\vec{k}} \right\rangle = H_{i\alpha,j\beta}^{\vec{k}}$, $\tilde{H}_{i'\alpha',j'\beta'}^{\vec{k}}$ is the matrix element of $\tilde{H}^{\vec{k}} = \left(S^{\vec{k}}\right)^{-1} H^{\vec{k}} \left(S^{\vec{k}}\right)^{-1}$, where

$S^{\vec{k}} = \left\langle \varphi_{i\alpha}^{\vec{k}} \middle| \varphi_{j\beta}^{\vec{k}} \right\rangle$ is the overlap matrix. The gradient of Eq. (4) to the atomic coordinate $\vec{\tau}_l$ is:

$$\frac{\partial \hat{H}^{\vec{k}}}{\partial \vec{\tau}_l} = \sum_{i'j'\alpha'\beta'} \left( \frac{\partial \tilde{H}_{i'\alpha',j'\beta'}^{\vec{k}}}{\partial \vec{\tau}_l} \left| \varphi_{i'\alpha'}^{\vec{k}} \right\rangle \left\langle \varphi_{j'\beta'}^{\vec{k}} \right| + \tilde{H}_{i'\alpha',j'\beta'}^{\vec{k}} \left| \frac{\partial \varphi_{i'\alpha'}^{\vec{k}}}{\partial \vec{\tau}_l} \right\rangle \left\langle \varphi_{j'\beta'}^{\vec{k}} \right| + \tilde{H}_{i'\alpha',j'\beta'}^{\vec{k}} \left| \varphi_{i'\alpha'}^{\vec{k}} \right\rangle \left\langle \frac{\partial \varphi_{j'\beta'}^{\vec{k}}}{\partial \vec{\tau}_l} \right| \right). \tag{5}$$

According to Eq. (5), the matrix element of the gradient of the Hamiltonian between the Bloch states $\left|\psi_{\mu}^{\vec{k}}\right\rangle = \sum_{i\alpha} C_{i\alpha}^{\vec{k}\mu} \left|\varphi_{i\alpha}^{\vec{k}}\right\rangle$ and $\left|\psi_{\upsilon}^{\vec{k}}\right\rangle = \sum_{j\beta} C_{j\beta}^{\vec{k}\nu} \left|\varphi_{j\beta}^{\vec{k}}\right\rangle$ is as follows:

$$\left\langle \psi_{\mu}^{\vec{k}} \middle| \frac{\partial \hat{H}^{\vec{k}}}{\partial \vec{\tau}_l} \middle| \psi_{\nu}^{\vec{k}} \right\rangle = \sum_{ii'jj'\alpha\alpha'\beta\beta'} C_{i\alpha}^{\vec{k}\mu *} C_{j\beta}^{\vec{k}\nu} \left( \frac{\partial \tilde{H}_{i'\alpha',j'\beta'}^{\vec{k}}}{\partial \vec{\tau}_l} S_{i\alpha,i'\alpha'}^{\vec{k}} S_{j'\beta',j\beta}^{\vec{k}} + \tilde{H}_{i'\alpha',j'\beta'}^{\vec{k}} \left\langle \varphi_{i\alpha}^{\vec{k}} \middle| \frac{\partial \varphi_{i'\alpha'}^{\vec{k}}}{\partial \vec{\tau}_l} \right\rangle S_{j'\beta',j\beta}^{\vec{k}} + \tilde{H}_{i'\alpha',j'\beta'}^{\vec{k}} S_{i\alpha,i'\alpha'}^{\vec{k}} \left\langle \frac{\partial \varphi_{j'\beta'}^{\vec{k}}}{\partial \vec{\tau}_l} \middle| \varphi_{j\beta}^{\vec{k}} \right\rangle \right). \tag{6}$$

Define $A_{i\alpha,i'\alpha'}^{\vec{k}l}$ and $A_{j\beta,j'\beta'}^{\vec{k}l*}$ as

$$A_{i\alpha,i'\alpha'}^{\vec{k}l} = \begin{cases} \left\langle \varphi_{i\alpha}^{\vec{k}} \middle| \frac{\partial \varphi_{i\alpha'}^{\vec{k}}}{\partial \vec{\tau}_i} \right\rangle & i' = l = i \\ 0 & i' = l \neq i \\ -\frac{\partial S_{i\alpha,i'\alpha'}^{\vec{k}}}{\partial \vec{\tau}_l} & i' \neq l \end{cases}, \quad A_{j\beta,j'\beta'}^{\vec{k}l*} = \begin{cases} \left\langle \varphi_{j\beta}^{\vec{k}} \middle| \frac{\partial \varphi_{j\beta'}^{\vec{k}}}{\partial \vec{\tau}_j} \right\rangle^* & j' = l = j \\ 0 & j' = l \neq j \\ -\frac{\partial S_{j\beta,j'\beta'}^{\vec{k}*}}{\partial \vec{\tau}_l} & j' \neq l \end{cases}. \tag{7}$$

Then the following two equations can be obtained

$$\left\langle \varphi_{i\alpha}^{\vec{k}} \middle| \frac{\partial \varphi_{i'\alpha'}^{\vec{k}}}{\partial \vec{\tau}_l} \right\rangle = \frac{\partial S_{i\alpha,i'\alpha'}^{\vec{k}}}{\partial \vec{\tau}_l} + A_{i\alpha,i'\alpha'}^{\vec{k}l}, \quad \left\langle \frac{\partial \varphi_{j'\beta'}^{\vec{k}}}{\partial \vec{\tau}_l} \middle| \varphi_{j\beta}^{\vec{k}} \right\rangle = \frac{\partial S_{j\beta,j'\beta'}^{\vec{k}*}}{\partial \vec{\tau}_l} + A_{j\beta,j'\beta'}^{\vec{k}l*}. \tag{8}$$

According to the derivative rules of matrices, the gradient of the matrix $\tilde{H} = S^{-1}HS^{-1}$ with respect to $\vec{\tau}_l$ is expanded as follows:

$$\frac{\partial \tilde{H}}{\partial \vec{\tau}_l} = -S^{-1}\frac{\partial S}{\partial \vec{\tau}_l}S^{-1}HS^{-1} + S^{-1}\frac{\partial H}{\partial \vec{\tau}_l}S^{-1} - S^{-1}HS^{-1}\frac{\partial S}{\partial \vec{\tau}_l}S^{-1} \tag{9}$$

Substitute Eqs. (8) and (9) into Eq. (6), then

$$\left\langle \psi_{\mu}^{\vec{k}} \middle| \frac{\partial \hat{H}^{\vec{k}}}{\partial \vec{\tau}_l} \middle| \psi_{\nu}^{\vec{k}} \right\rangle = \sum_{i\alpha,j\beta} C_{i\alpha}^{\vec{k}\mu *} C_{j\beta}^{\vec{k}\nu} \left[ \frac{\partial H^{\vec{k}}}{\partial \vec{\tau}_l} + \left( A^{\vec{k}l} \left( S^{\vec{k}} \right)^{-1} H^{\vec{k}} \right) + \left( H^{\vec{k}} \left( S^{\vec{k}} \right)^{-1} \left( A^{\vec{k}l} \right)^* \right) \right]_{i\alpha,j\beta} \tag{10}$$

$A^{\vec{k}l}$ and $S^{\vec{k}}$ are known because they can be calculated analytically in the atomic basis functions. The Hamiltonian matrix $H^{\vec{k}}$ can be predicted directly through the forward propagation of the networks, and the gradient $\frac{\partial H^{\vec{k}}}{\partial \vec{\tau}_l}$ of the Hamiltonian matrix elements can be obtained through the backward propagation or the finite difference of the predicted Hamiltonian matrix. Therefore, the EPC matrix at Γ point can be predicted by Eqs. (10) and (3).

When $\mu \neq \upsilon$, $\left\langle \psi_\mu^{\vec{k}} \middle| \frac{\partial \hat{H}^{\vec{k}}}{\partial \vec{\tau}_l} \middle| \psi_\nu^{\vec{k}} \right\rangle$ can be determined by the following equation[24, 25]:

$$\left\langle \psi_\mu^{\vec{k}} \middle| \frac{\partial \hat{H}^{\vec{k}}}{\partial \vec{\tau}_l} \middle| \psi_\upsilon^{\vec{k}} \right\rangle = \left( \varepsilon_\upsilon^{\vec{k}} - \varepsilon_\mu^{\vec{k}} \right) \vec{d}_{\mu\nu,l}^{(\vec{k})}, \quad \vec{d}_{\mu\nu,l}^{(\vec{k})} = \left\langle \psi_\mu^{\vec{k}} \middle| \frac{\partial \psi_\upsilon^{\vec{k}}}{\partial \vec{\tau}_l} \right\rangle, \tag{11}$$

where $\vec{d}_{\mu\nu,l}^{(\vec{k})}$ is called the non-adiabatic coupling vector (NACV). In the following section, we only use Eq. (11) to verify the correctness of Eq. (10) and compare it with the analytic gradient from ML.

### *The actual tests*

The accurate calculation of the phonon spectrum is a mature technique and can be done with the famous phonopy[26] package. Here we assume that we already have the eigenvector $\xi_{l_\sigma}^{b\vec{q}}$ of each phonon mode. We only test the accuracy of the predicted values of Eq. (10) at Γ point. The tests were carried out on a water molecule and a 2×2×1 supercell of $MoS_2$ crystal.

We first trained two HamGNN models on the Hamiltonian datasets of 500 water molecules and 500 perturbed $MoS_2$ structures, respectively. A water molecule and a $MoS_2$ crystal were then chosen from the test set respectively to test the accuracy of the predicted expectation value of the Hamiltonian gradient operator. We first compared the difference between the DFT calculated $\frac{\partial H_{i\alpha,j\beta}^{\vec{k}}}{\partial \vec{\tau}_l}$ using the finite difference method and the value predicted by HamGNN. As shown in Fig. 2, the gradients of the Hamiltonian matrix elements calculated by DFT and predicted by HamGNN are in good agreement. The mean absolute errors (MAEs) of the predicted $\frac{\partial H_{i\alpha,j\beta}^{\vec{k}}}{\partial \vec{\tau}_l}$ at the Γ point for the testing water molecule and $MoS_2$ crystal are within ~$10^{-5}$ Hatree/Bohr. Table 1 and Table 2 list the DFT calculated and HamGNN predicted values of $\left\langle \psi_\mu \middle| \frac{\partial \hat{H}}{\partial \vec{\tau}_l} \middle| \psi_\upsilon \right\rangle$ for the water molecule and $MoS_2$ crystal, respectively. Except for the matrix elements below $10^{-3}$ that can be ignored, the values predicted by the HamGNN network are in good agreement with those calculated by DFT using Eqs. (10) and (11), indicating that Eq. (10) is correct and that the prediction accuracy is comparable to DFT.

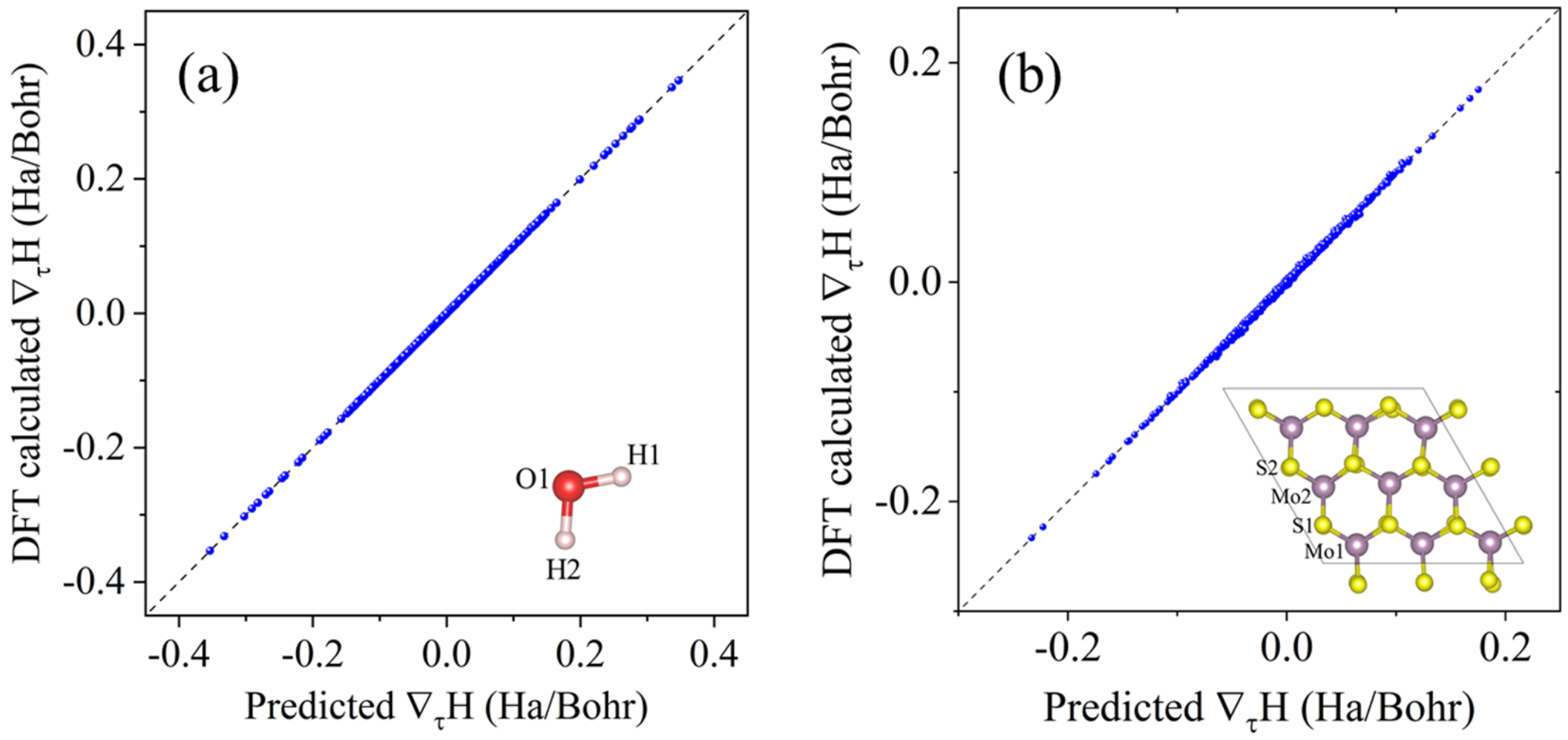

**Fig. 2**. Comparison of the calculated and predicted derivatives of the Hamiltonian elements at the Γ point for the testing (a) water molecule and (b) $MoS_2$ crystal.

**Table 1**. The DFT calculated and HamGNN predicted $\left\langle \psi_\mu \middle| \frac{\partial \hat{H}}{\partial \vec{\tau}_l} \middle| \psi_\upsilon \right\rangle$ for the testing water molecule (eV/Å).

| $\mu$ | $N$ | Atoms | Calculated EPC[a] | Calculated EPC[b] | Predicted EPC |
|---|---|---|---|---|---|
| LUMO | LUMO+1 | H1 | -5.52<br>-1.13<br>$-9.60\times10^{-3}$ | -5.53<br>-1.13<br>$-1.17\times10^{-2}$ | -5.52<br>-1.14<br>$-1.23\times10^{-2}$ |
| | | H2 | -0.48<br>-5.60<br>$8.87\times10^{-3}$ | -0.49<br>-5.60<br>$1.10\times10^{-2}$ | -0.49<br>-5.61<br>$8.73\times10^{-3}$ |
| | | O1 | 5.64<br>6.32<br>$-1.23\times10^{-3}$ | 5.64<br>6.31<br>$-3.83\times10^{-5}$ | 5.64<br>6.32<br>$-1.13\times10^{-3}$ |

[a]calculated by Eq. (10).
[b]calculated by Eq. (11).

**Table 2**. The DFT calculated and HamGNN predicted $\left\langle \psi_\mu \middle| \frac{\partial \hat{H}}{\partial \vec{\tau}_l} \middle| \psi_\upsilon \right\rangle$ for the testing $MoS_2$ crystal (eV/Å). Only the gradients to the coordinates of Mo1, Mo2, S1, and S2 are listed in the table.

| $\mu$ | $N$ | Atoms | Calculated EPC[a] | Calculated EPC[b] | Predicted EPC |
|---|---|---|---|---|---|
| LUMO | LUMO+1 | Mo1 | $-4.78\times10^{-2}$<br>$-8.78\times10^{-2}$<br>$-6.28\times10^{-2}$ | $-4.76\times10^{-2}$<br>$-9.09\times10^{-2}$<br>$-6.44\times10^{-2}$ | $-5.10\times10^{-2}$<br>$-8.83\times10^{-2}$<br>$-6.80\times10^{-2}$ |
| | | Mo2 | $9.16\times10^{-2}$<br>$1.50\times10^{-1}$<br>$2.62\times10^{-2}$ | $9.03\times10^{-2}$<br>$1.50\times10^{-1}$<br>$2.78\times10^{-2}$ | $9.71\times10^{-2}$<br>$1.57\times10^{-1}$<br>$2.44\times10^{-2}$ |
| | | S1 | $-6.18\times10^{-2}$<br>$4.34\times10^{-2}$<br>$3.42\times10^{-3}$ | $-6.19\times10^{-2}$<br>$4.13\times10^{-2}$<br>$2.43\times10^{-3}$ | $-5.35\times10^{-2}$<br>$3.11\times10^{-2}$<br>$6.43\times10^{-3}$ |
| | | S2 | $3.28\times10^{-2}$<br>$3.04\times10^{-2}$<br>$9.57\times10^{-3}$ | $3.05\times10^{-2}$<br>$2.99\times10^{-2}$<br>$9.55\times10^{-3}$ | $2.00\times10^{-2}$<br>$2.37\times10^{-2}$<br>$1.53\times10^{-2}$ |

[a]calculated by Eq. (10).
[b]calculated by Eq. (11).

## Discussion

In this work, we derived an EPC calculation formula in the non-orthogonal atomic orbital basis, which can be expressed as a function of the Hamiltonian matrix and its derivatives. The latter two matrices can be obtained by the forward propagation and backward propagation of the E(3) equivariant HamGNN model, respectively. By testing on a water molecule and a $MoS_2$ crystal, we verified the correctness of the proposed EPC formula and the high accuracy of the values predicted by the trained models. Our proposed method provides a shortcut to accelerate the calculation of the EPC matrix.

## Acknowledgment

We thank Prof. Honghui Shang from the Chinese Academy of Sciences for the useful discussions. We acknowledge financial support from the Ministry of Science and Technology of the People´s Republic of China (No. 2022YFA1402901), NSFC (grants No. 11825403, 11991061, 12188101), and the Guangdong Major Project of the Basic and Applied Basic Research (Future functional materials under extreme conditions--2021B0301030005).